\documentclass[aps,twocolumn,groupedaddress,showpacs,nofootinbib]{revtex4}
\usepackage{epsfig,amsbsy,bm,amssymb,amsmath}
\usepackage{graphics}

\usepackage{amsmath}  
\usepackage{amsfonts} 
\usepackage{graphicx} 

\newcommand{\hs}{\hspace*}
\newcommand{\vs}{\vspace*}

\newcommand{\eref}[1] {(\ref{#1})}
\newcommand{\Eref}[1] {Eq.~(\ref{#1})}
\newcommand{\Fref}[1] {Fig. \ref{#1}}

\newcommand{\nn}{\nonumber}
\newcommand{\be}{\begin{equation}}
\newcommand{\ee}{\end{equation}}
\newcommand{\br}{\begin{eqnarray*}}
\newcommand{\er}{\end{eqnarray*}}
\newcommand{\ba}{\begin{eqnarray}}
\newcommand{\ea}{\end{eqnarray}}
\newcommand{\bp}{\begin{minipage}}
\newcommand{\ep}{\end{minipage}}

\newcommand{\e}{\epsilon}
\renewcommand{\l}{\lambda}

\newcommand{\three}[6]
{\left(\!\!\!\begin{array}{rrr}
{#1\!}&{#2\!}&{#3\!}\\
{#4\!}&{#5\!}&{#6\!}\\
\end{array}\!\right)}
\begin{document}

\title{Relativistic calculations of angular dependent photoemission
time delay}

\author{Anatoli Kheifets}
\email{a.kheifets@anu.edu.au}

\affiliation{Research School of Physics and Engineering, The
Australian National University, Canberra ACT 0200, Australia}

\author{Ankur Mandal}
\email{amankur@physics.iitm.ac.in} 

\affiliation{Department of Physics, Indian Institute of Technology
Madras, Chennai, Tamil Nadu 600036, India}

\author{Pranawa C. Deshmukh}
\email{pcd@physics.iitm.ac.in}

\affiliation{Department of Physics, Indian Institute of Technology
Madras, Chennai, Tamil Nadu 600036, India}

\author{Valeriy K. Dolmatov}
\email{vkdolmatov@una.edu}

\affiliation{Department of Physics and Earth Science, University of
North Alabama, Florence, AL 35632, USA }

\author{David A. Keating}
\email{dkeating2@student.gsu.edu}

\affiliation{Department of Physics and Astronomy, Georgia State
University, Atlanta, GA 30303, USA}

\author{Steven T. Manson}
\email{smanson@gsu.edu}

\affiliation{Department of Physics and Astronomy, Georgia State
University, Atlanta, GA 30303, USA}


\date{\today}

\begin{abstract}
Angular dependence of photoemission time delay for the valence
$np_{3/2}$ and $np_{1/2}$ subshells of Ar, Kr and Xe is studied in the
dipole relativistic random phase approximation. Strong angular
anisotropy of the time delay is reproduced near respective Cooper minima
while the spin-orbit splitting affects the time delay near threshold.
\end{abstract}


\maketitle 

\section{Introduction} 

A measurable time delay in laser driven atomic ionization has been
discovered recently \cite{M.Schultze06252010,PhysRevLett.106.143002}.
Since the first pioneering experiments, the time-delay spectroscopy of
laser-induced atomic ionization (attosecond chronoscopy) has become a
rapidly developing field \cite{RevModPhys.87.765}.  Among other
characteristic features, an angular anisotropy of attosecond time
delay relative to polarization of laser light has been predicted
theoretically \cite{0953-4075-48-2-025602,Dahlstrom2014} and measured
experimentally \cite{2015arXiv150308966H}.  In one-photon
photoionization of atomic $np$ subshells, the time delay can show some
angular anisotropy due to the interplay of the $\epsilon s$ and
$\epsilon d$ photoelectron continua
\cite{0953-4075-48-2-025602,Dahlstrom2014}. This anisotropy becomes
particularly strong near a Cooper minimum in the dominant $np\to\e d$
channel, making the nominally weak $np\to\e s$ channel competitive. In
two-color (two-photon) XUV/IR experiments, the interference of these
photoemission channels can manifest itself even in a spherically
symmetric $ns$ atomic subshell. This leads to a strong angular
anisotropy of the measured time delay when the $\epsilon d$ continuum
has a kinematic node near the magic angle of $54.7^\circ$. Such a
strong anisotropy has indeed been measured in He in a recent RABBITT
(Reconstruction of Attosecond Beating By Interference of Two-photon
Transitions) experiment\cite{2015arXiv150308966H}. Another interesting
aspect of photoemission time delay is its sensitivity to the fine
structure of the ionized target. Recent experiments have detected such
a sensitivity in the valence shell photoionization of the Kr and Xe
atoms \cite{1742-6596-635-9-092135}.

In this paper, we investigate both the angular and spin dependence of
the time delay using the dipole relativistic random phase
approximation (RRPA). We expand our previous relativistic studies of
the time delay \cite{PhysRevA.90.053406,PhysRevA.92.063422} and
include the full interference of all the spin-orbit coupled
photoionization channels. In the previous studies, only the time delay
in the dominant channel was evaluated. We validate our theoretical
model using the angular dependent time delay near the Cooper minimum
of the $3p$ subshell of Ar. For the relatively light Ar atom, our RRPA
results agree very well with non-relativistic random-phase
approximation with exchange (RPAE) \footnote{The same
  exchange interaction is accounted for both in the RPAE and RRPA but
  E is dropped from the latter acronym for brevity.}
calculations \cite{0953-4075-48-2-025602}. For heavier Kr and Xe
atoms, we clearly observe the manifestation of relativistic
effects. One such effect is a spin-orbit splitting of the time delay
near threshold.

The paper is organized as follows. In Sec.~\ref{Theory}  a a brief theoretical
formulation is given.  In section III the results for the angle and
energy dependence of Wigner time delay for photoemission from outer
$np_{1/2}$ and $np_{3/2}$ subshells of atomic Ar, Kr and Xe are
presented and discussed; Ne is omitted because there is no Cooper minimum 
in its photoionization cross section.   Conclusions are drawn in section IV.

\section{Theoretical Method}
\label{Theory}

\subsection{Photoionization amplitude}

We adopt the multichannel RRPA formalism of \citet{PhysRevA.20.964}.
In this formalism, the amplitude for a transition from the
ground state ($u_{i}$) to an excited state ($\omega_{i\pm}$), induced
by a time varying external field $v_+e^{-i\omega t} + v_-e^{i\omega
t}$ is given by
\be 
T=\sum_{i=1}^{N}  \int d^{3}r(\omega_{i+}^{\dagger}\vec{\alpha}\cdot\vec{A}
u_{i}+u_{i}^{\dagger} \vec{\alpha}\cdot\vec{A}\omega_{i-})
\ .  
\ee
Here the electromagnetic interaction is written in Coulomb gauge
and expressed in terms of the Pauli spin matrices 
$\vec{\alpha} =\left(
\begin{array}{c c} 0 & \vec{\sigma}\\ \vec{\sigma} & 0 \end{array}
\right)$
and the vector potential $\vec{A}$.

In a single active electron approximation, the multipole transition
amplitude is reduced to 
\be 
T_{JM}^{(\lambda)}=\int
d^{3}r{\omega_{i+}^{\dagger}\vec{\alpha}\cdot\vec{a}_{JM}^{\lambda}u_{i}}
\ ,
\ee
where the indices $J$ and $M$ are the photon angular momentum and its
projection and $\lambda = 1$ or 0 for electric or magnetic multipoles,
respectively.  Specifically, for a one-electron transition from an
initial state characterized by the quantum numbers $ ljm$ to a final
continuum state $\bar{l}\bar{j}\bar{m}$ with the spin described by
a two-component spinor $\chi_{\nu}$, this equation becomes 
\ba 
T_{JM}^{(\lambda)}&=&i {\sqrt{\frac{2\pi^{2}}{Ep}}}
{\sqrt{\frac{(2J+1)(J+1)}{J}}} {\frac{\omega^{J}}{(2J+1)!!}}  
\nn\\&&\hs{-1cm}\times
\sum_{\bar{\kappa}\bar{m}}(\chi_{\nu}^{\dagger}
\Omega_{\bar{\kappa}\bar{m}}(\hat{p})) (-1)^{\bar{j}-\bar{m}}
\left(
\begin{array}{c c c} \bar{j} & J & j \\ -\bar{m} & M & m \end{array}
\right)
\nn\\&& \times
i^{1-\bar{l}} e^{i\delta_{\bar{\kappa}}} \left\langle \bar{a}
\|Q_J^{(\lambda)} \|a\right\rangle
(-1)^{\bar j+ j +J}
\ .
\label{TJM}
\ea

\noindent
Here $E$ and $\hat{p}$ are the photoelectron energy and momentum
direction, respectively, $\omega$ is the photon frequency, 
$\delta_{\bar{\kappa}}$ is the phase of the continuum wave with
$\bar{\kappa} = \mp(\bar{j}+\frac{1}{2})$ for $\bar{j} = (\bar{l} \pm
\frac{1}{2})$.  The spherical spinor is defined as 
\be
\Omega_{\kappa m}(\hat{n}) = 
\sum_{\nu=\pm1/2}
C^{jM}_{l,M-\nu,1/2\nu}
Y_{l m-\nu} (\hat{n})\chi_{\nu} 
\ee
The corresponding Clebsch-Gordan coefficients are tabulated in
\cite{V88}.
The reduced matrix element of the spherical tensor between the initial
sate $a = (n\kappa)$ and a final energy scale normalized state $a= (E,
\bar\kappa)$ is written as
\ba
\left\langle \bar{a} \|Q_J^{(\lambda)}\|a\right\rangle
&=&
(-1)^{j+1/2}[\bar j][j]
\three{j}{\bar j} {J}
{-1/2} {1/2}{0}
\nn\\&&\times \
\pi(\bar l,l,J-\l+1)
R^{(\l)}_J(\bar a,a)
\label{reduced}
\ea
Here $\pi(\bar l,l,J-\l+1)=1$ or 0 for 
$\bar l+l+J-\l+1$ even or odd, respectively, and $R^{(\l)}_J(\bar a,a)$
is the radial integral. While \Eref{reduced} is derived for a
single-electron transition, it also applies to closed-shell atomic
systems.  In order to include the RRPA correlations, the only change
in \Eref{TJM} is to replace $\left\langle \bar{a} \|Q_J^{(\lambda)}
\|a\right\rangle$ with $\left\langle \bar{a} \|Q_J^{(\lambda)}
\|a\right\rangle_{\rm RRPA}$.  Finally, as we will be dealing with
electric dipole photoionizing transitions, we set $\l=1$, $J=1$ and
choose $M=0$ which corresponds to linear polarization in the
$z$-direction.  In this case, \Eref{TJM} is reduced to
\ba
T_{1 0}^{1\pm} \equiv [T_{1 0}^{(1)}]_{\nu = \pm1/2}
&=& 
\sum_{\bar{\kappa}\bar{m}}
C^{jM}_{l,M-\nu,1/2\nu}
Y_{l m-\nu} (\hat p)\chi_{\nu} 
\nn 
\\&&\hs{-4cm}\times
(-1)^{\bar{j}+j+1+ \bar{j}-\bar{m}}
\left(
\begin{array}{c c c} \bar{j} & 1 & j \\ -\bar{m} & 0 & m \end{array}
\right) 
i^{1-\bar{l}} e^{i\delta_{\bar{\kappa}}} \left\langle \bar{a}
\|Q_1^{(1)} \|a\right\rangle
\ .
\label{ampl}
\ea
Here we dropped the common scaling factor for brevity of
notation.  We will also be using a shorthand for a reduced matrix
element modified by the phase factors:
\be 
D_{lj \to \bar{l}\bar{j}}= i^{1-\bar{l}} e^{i\delta_{\bar{\kappa}}}
\left\langle \bar{a} \|Q_J^{(\lambda)} \|a\right\rangle
\label{phase_factor}
\ee
We note that \Eref{TJM} differs by the extra parity factor $(-1)^{\bar
  j+ j +J}$ from the original equation~(43) of
\citet{PhysRevA.20.964}. We added this factor to make it comply with
the Wigner-Eckart theorem (Eq.~107-6 of \citet{LL85}).

\subsection{Formulation of angular dependent time delay}

An electric dipole transition from a $np$
initial state leads to the following five ionization channels:
\br
np_{1/2} &\to& \e s_{1/2}, \ \e d_{3/2} \\
np_{3/2} &\to& \e s_{1/2}, \ \e d_{3/2} , \ \e d_{5/2} \\
\er
Using \Eref{ampl}, we derive the following expressions for the
$np_{1/2}$ ionization amplitude:
\br
{[T_{1 0}^{1+}]}_{np_{1/2}}^{m=\frac12}
&=& 
+ {1\over\sqrt{15}} Y_{20}D_{np_{1/2}\to\e d_{3/2}}
+ {1\over\sqrt6} Y_{00} D_{np_{1/2}\to\e s_{1/2}}
\\
{[T_{1 0}^{1-}]}_{np_{1/2}}^{m=\frac12}
&=& 
-{1\over\sqrt{10}} Y_{21} \ D_{np_{1/2}\to\e d_{3/2}}
\er
Here and throughout the text, $Y_{l,m} \equiv Y_{lm}(\hat p)$. The
corresponding amplitudes with the $m=-1/2$ projection will have a
similar structure with the simultaneous inversion of the spin
projection $T^+\leftrightarrow T^-$ and the second index of the
spherical harmonic $Y_{21}\to Y_{2-1}$.
Each amplitude has its own associated photoelectron group delay (the
Wigner time delay \cite{PhysRev.98.145}) defined as
\be
\tau =  \frac{d\eta}{dE}
\ \ , \ \ 
\eta = \tan^{-1}\left[
{{\rm Im} T_{1 0}^{1\pm} \over
{\rm Re} T_{1 0}^{1\pm}}
\right]
\ .
\ee
The spin averaged time delay can be expressed as a weighted sum
\be
\bar\tau_{np_{1/2}} = {
\tau^{m=\frac12,+}_{np_{1/2}} 
\left|[T^{1+}_{10}]_{np_{1/2}}^{m=\frac12}\right|^2
+
\left|[T^{1-}_{10}]_{np_{1/2}}^{m=\frac12}\right|^2
\tau^{m=\frac12,-}_{np_{1/2}} 
\over
\left|[T^{1+}_{10}]_{np_{1/2}}^{m=\frac12}\right|^2
+
\left|[T^{1-}_{10}]_{np_{1/2}}^{m=\frac12}\right|^2
}
\ee
The angular resolved amplitudes for the $np_{3/2}$ initial state take
the following form:
\br
{[T_{1 0}^{1+}]}_{np_{3/2}}^{m=1/2}
&=& 
 {1\over\sqrt6} Y_{00} D_{np_{3/2}\to\e s_{1/2}}
- {1\over5\sqrt{6}}Y_{20} D_{np_{3/2}\to\e d_{3/2}}
\\&&
- {1\over5}\sqrt{\frac32}Y_{20} D_{np_{3/2}\to\e d_{5/2}}
\\
{[T_{1 0}^{1-}]}_{np_{3/2}}^{m=1/2}
&=& 
{1\over10} Y_{21}D_{np_{3/2}\to\e d_{3/2}}
- {1\over5} Y_{21}D_{np_{3/2}\to\e d_{5/2}}
\\
{[T_{1 0}^{1+}]}_{np_{3/2}}^{m=3/2} 
&=& 
-{\sqrt3\over 10} Y_{21}D_{np_{3/2}\to\e d_{3/2}}
-{2\sqrt3\over 15} Y_{21}D_{np_{3/2}\to\e d_{5/2}}
\\
{[T_{1 0}^{1-}]}_{np_{3/2}}^{m=3/2}
&=& 
{\sqrt3\over 5} Y_{22}D_{np_{3/2}\to\e d_{3/2}}
-{\sqrt3\over 15} Y_{22}D_{np_{3/2}\to\e d_{5/2}}
\er
The corresponding spin averaged time delay becomes
\br
\hs{-0.5cm}
{\cal S} \times \bar\tau_{np_{3/2}} &=& 
\tau^{m=\frac12,+}_{np_{3/2}} 
\left|[T^{1+}_{10}]_{np_{3/2}}^{m=\frac12}\right|^2
+
\tau^{m=\frac12,-}_{np_{3/2}} 
\left|[T^{1-}_{10}]_{np_{3/2}}^{m=\frac12}\right|^2
\\&+&
\tau^{m=\frac32,+}_{np_{3/2}} 
\left|[T^{1+}_{10}]_{np_{3/2}}^{m=\frac32}\right|^2
+
\tau^{m=\frac32,-}_{np_{3/2}} 
\left|[T^{1-}_{10}]_{np_{3/2}}^{m=\frac32}\right|^2
\\
{\rm with} \ \ \ \   {\cal S} &=&  
\left|[T^{1+}_{10}]_{np_{3/2}}^{m=\frac12}\right|^2
+
\left|[T^{1-}_{10}]_{np_{3/2}}^{m=\frac12}\right|^2
\\&&\hs{1cm}+
\left|[T^{1+}_{10}]_{np_{3/2}}^{m=\frac32}\right|^2
+
\left|[T^{1-}_{10}]_{np_{3/2}}^{m=\frac32}\right|^2
\er
Here we restricted ourselves to positive $m=1/2,3/2$. The
corresponding time delay with negative $m$ will be identical.

\subsection{Non-relativistic limit}

Using Eqs.~\eref{reduced} and \eref{phase_factor}, we can express the
amplitudes via the corresponding radial integrals modified by the
phase factors. For a mildly
relativistic atom, the radial integrals with $\bar j=\bar l\pm1/2$
orbitals will be very similar. Hence we can neglect this difference
and reduce the amplitudes to the following expressions:
\br
T_{1 0}^{1+} (np_{1/2})
&=& 
- \frac13 Y_{00}  R_{np\to \e s} 
- {2\over3} {1\over\sqrt5} Y_{20}  R_{np\to \e d}
\\
T_{1 0}^{1+} (np_{3/2})
&=& 
 {\sqrt2\over3} Y_{00} R_{np\to \e s}
+ 
{2\over3}\sqrt{\frac25}
Y_{20} R_{\e d} \ \ 
\\
T_{1 0}^{1-} (np_{1/2})
&=& 
\sqrt{2\over15} Y_{21}  R_{np\to \e d}
\\
T_{1 0}^{1-}(np_{3/2})  &=& 
 \sqrt{1\over15} Y_{21} R_{np\to \e d}
\er
This is to be compared with the corresponding non-relativistic amplitudes
\cite{PhysRevA.87.063404}
\br
T_{np_{m=0}\to \e s}&=&
{1\over\sqrt3} \
Y_{00}(\hat n)\
R_{np\to \e s} 
\\
T_{np_{m=0}\to \e d}&=&
2\sqrt{1\over15} \
Y_{20}(\hat n)\
R_{np\to \e d} 
\\
T_{np_{m=1}\to Ed} &=&
-\sqrt{1\over5} \
Y_{21}(\hat n)\
R_{np\to Ed} 
\er
%

By comparing the weakly relativistic and strictly non-relativistic
amplitudes, we can observe the following scaling properties:
\ba
{[T^+]}_{np_{1/2}}^{m=1/2}
\simeq {1\over\sqrt3} T_{np_{m=0}}
& , &
{[T^-]}_{np_{1/2}}^{m=1/2}
\simeq -\sqrt\frac23 T_{np_{m=1}}
\nn\\
{[T^+]}_{np_{3/2}}^{m=1/2}
\simeq \sqrt\frac23 T_{np_{m=0}}
& , &
{[T^-]}_{np_{3/2}}^{m=1/2}
\simeq -{1\over\sqrt3} T_{np_{m=1}}
\nn\\
{[T^+]}_{np_{3/2}}^{m=3/2}\simeq T_{np_{m=1}}
& , &
{[T^-]}_{np_{3/2}}^{m=3/2}\simeq 0
\label{scaling}
\ea
By feeding this scaling into the spin-averaged time delay expressions,
we get
\br
\bar \tau_{np_{1/2}} &=& 
{
\tau_{np_{m=0}} 
|T_{np_{m=0}}|^2
+2
\tau_{np_{m=1}} 
|T_{np_{m=1}}|^2
\over
|T_{np_{m=0}}|^2
+
2
|T_{np_{m=1}}|^2
}
\simeq \bar \tau_{np} \ ,
\er
which represents  the magnetic projection average nonrelativistic time
delay as is used in \cite{0953-4075-48-2-025602}. By the same token,
\br
\bar\tau_{np_{3/2}} &=& 
{ 
2\tau_{np_{m=0}} |T_{np_{m=0}}|^2
+
4\tau_{np_{m=1}} |T_{np_{m=1}}|^2
\over
2 |T_{np_{m=0}}|^2
+
4 |T_{np_{m=1}}|^2
}
\simeq\bar\tau_{np}
\er

\section{Results and Discussion}

The RRPA calculations have been performed with the following channels
coupled: 14 channels for Ar (all dipole excitations from the $3p$,
$3s$ and $2p$ subshells), 18 channels for Kr (excitations from the $4p$,
$4s$, $3d$ and $3p$ subshells) and 18 channels for Xe (the $5p$, $5s$,
$4d$ and $4p$ subshells).  In addition, where available, experimental
threshold energies from \cite{NIST-ASD} were used to facilitate
better comparison with experiment. These energies are identical to
those displayed in Table~1 of \cite{PhysRevA.90.053406}.

\subsection{Argon}

\begin{figure}[ht]

\epsfxsize=0.9\columnwidth
\epsffile{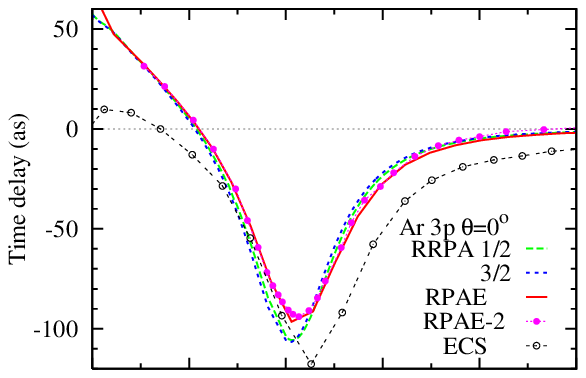}
\vs{-7mm}

\epsfxsize=0.9\columnwidth
\epsffile{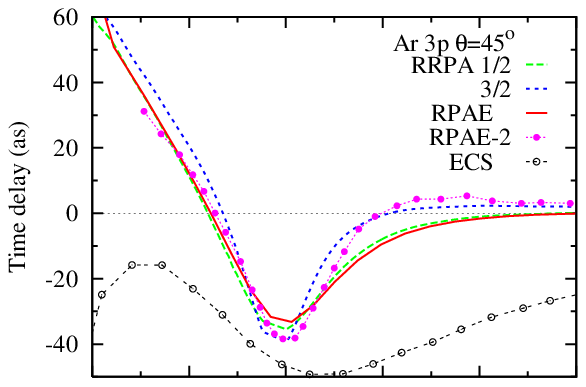}
\vs{-7mm}

\epsfxsize=0.9\columnwidth
\epsffile{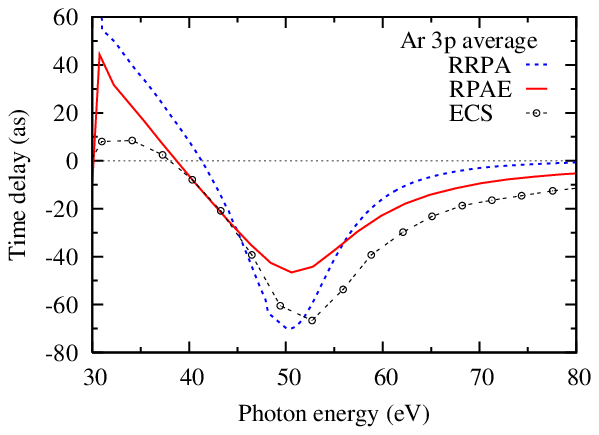}

\caption{(Color online) The time delay of Ar $3p_{1/2}$ (green dashed
line) and $3p_{3/2}$ (blue dotted line) is plotted versus the photon
energy.  The nonrelativistic RPAE results from
Ref.~\cite{PhysRevA.87.063404} and RPAE-2 from
\cite{0953-4075-48-2-025602} are shown by the red solid line and
filled circles, respectively. The top panel: $\theta=0^\circ$, middle panel:
$\theta=45^\circ$, bottom panel: angular average. The experimental
data from \cite{0953-4075-47-24-245003} are shown on the bottom panel
with error bars.}
\label{Fig1}
\end{figure}

Argon is hardly a relativistic target and only chosen here for
validation and calibration of our theoretical model against the
previous calculations \cite{0953-4075-48-2-025602,Dahlstrom2014} and
experiment \cite{0953-4075-47-24-245003}.
The time delay of Ar $3p_{1/2}$ and $3p_{3/2}$ subshells in the energy
range encompassing the Cooper minimum is shown in \Fref{Fig1} for the
fixed emission angles $\theta=0^\circ$ (top panel) and $\theta=45^\circ$
(middle panel). Comparison is made with the non-relativistic RPAE
calculations \cite{PhysRevA.87.063404,0953-4075-48-2-025602} and the
exterior complex scaling (ECS) calculation \cite{Dahlstrom2014}. 
The bottom panel displays the angular averaged result where a
comparison is also shown with a RABBITT experiment
\cite{0953-4075-47-24-245003} in which no discrimination with
the photoelectron direction was made. 

In the polarization direction at $\theta=0^\circ$, the present RRPA
results and non-relativistic RPAE
\cite{PhysRevA.87.063404,0953-4075-48-2-025602} results agree very
well. Not surprisingly, there is very little difference in the time
delay between the $3p_{1/2}$ and $3p_{3/2}$ subshells in Ar,
indicating that the spin-orbit interaction is unimportant here .  The
ECS calculation \cite{Dahlstrom2014} differs noticeably. This may be
attributed to a different methodology of the time delay determination
in this work. Indeed, all the RRPA and RPAE calculations were
performed for single-photon XUV photoionization. Hence it is the
Wigner time delay that was evaluated in the present work. In the ECS
calculation \cite{Dahlstrom2014}, the time delay was computed using
correlated two-photon (XUV+IR) above threshold matrix elements
\cite{PhysRevA.86.061402}. In such a definition, the atomic time delay
contains the so-called Coulomb-laser coupling (CLC) correction.
\be
\tau_a = \tau_{\rm W} + \tau_{\rm CLC}
\ ,
\label{atomic}
\ee
This correction is known to decrease rapidly with a growing
photoelectron energy and is relatively small near the Cooper
minimum. However, it is large near threshold and will be accounted for
in our calculations presented in the next section.
At the fixed photoelectron emission angle $\theta=45^\circ$ (middle
panel), the RPAE results
\cite{PhysRevA.87.063404,0953-4075-48-2-025602} and the present RRPA
calculation are still rather close. However, the ECS calculations is
considerably further away than in the case of $\theta=0^\circ$. This
deviation can be possibly attributed to further suppression of the $\e
d$ continuum because of the kinematic node of the $f$-wave close to
$45^\circ$. We note that the $\e d$ continuum is converted to the 
$p$- and $f$-waves by the IR photon absorption in two-color
photoionization experiments.  
When the angular average is taken (bottom panel), the difference
between all the calculations is not so pronounced and they compare
reasonably well with the experiment \cite{0953-4075-47-24-245003}.  A
more accurate angular resolved measurement is needed to validate various
theoretical predictions  in the directions away from the polarization
axis.

\subsection{Krypton and xenon}

\begin{figure}[ht]

\epsfxsize=0.9\columnwidth
\epsffile{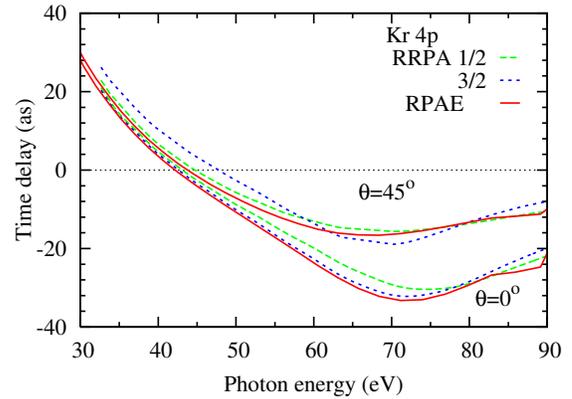}

\caption{ The time delay of Kr $4p_{1/2}$ (green dashed line) and
$4p_{3/2}$ (blue dotted line) is plotted versus the photon energy in
the fixed emission directions $\theta=0^\circ$ and $\theta=45^\circ$.  The
nonrelativistic RPAE result from Ref.~\cite{PhysRevA.87.063404} is
shown by the red solid line.  The angular averaged experimental data
from \cite{0953-4075-47-24-245003} are shown with error bars.
\label{Fig2}
}
\end{figure}

The time delay of Kr $4p_{1/2}$ and $4p_{3/2}$ subshells in the energy
range encompassing the Cooper minimum is shown in \Fref{Fig2} in the
two fixed directions $\theta=0^\circ$ and $\theta=45^\circ$. The time
delay displays the characteristic dip near the Cooper minimum but not
as deep as in the case of argon. The depth of the minimum indicates
the relative strength of the nominally weaker channel near the Cooper
minimum of the normally stronger channel.  In the non-relativistic
RPAE model, this stronger channel can be identified with the
$4p_{m=0}\to \epsilon d$ transition. Other angular momentum
projections $m=\pm1$ in this channel are excluded in the
$\theta=0^\circ$ polarization direction. Away from this direction,
weaker channels $4p_{m=\pm1}\to \epsilon d$, along with the
$4p_{m=0}\to \epsilon s$ channel, also make their contribution to the
photoionization amplitude and the time delay. Hence the Cooper minimum
in the time delay is getting shallower.  Overall the angular
dependence is much weaker in Kr than in the case of Ar.  Nevertheless,
the RRPA calculations show a noticeable deviation from the RPAE and
the spin-orbit splitting of the time delay becomes visible.

\begin{figure}[ht]

\epsfxsize=0.9\columnwidth
\epsffile{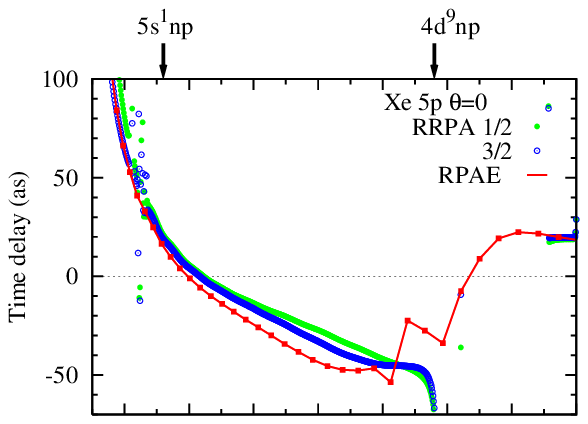}
\vs{-7.5mm}

\epsfxsize=0.9\columnwidth
\epsffile{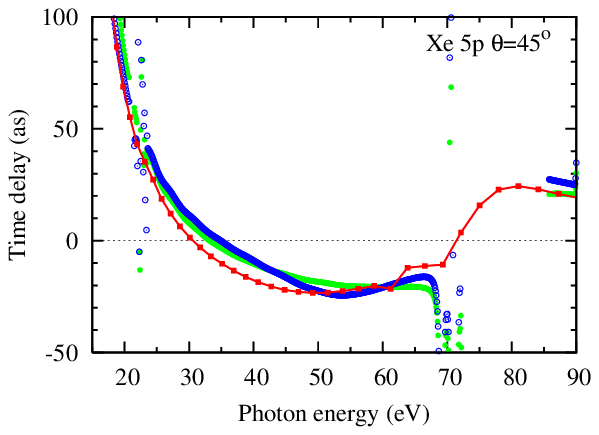}

\caption{ The time delay of Xe $5p_{1/2}$ (green filled circles) and
$5p_{3/2}$ (blue open circles) is plotted versus the photon energy.
The nonrelativistic RPAE result from Ref.~\cite{PhysRevA.87.063404} is
shown by the red solid line. Regions of the autoionization
series $5s^1np$ \cite{PhysRevA.4.2261} and $4d^9np$ \cite{Ederer:75}
are marked by vertical arrows.
\label{Fig3}
}
\end{figure}

The analogous set of data for the Xe $5p_{1/2}$ and $5p_{3/2}$ subshells
is shown in \Fref{Fig3}. The photon energy range near the Cooper
minimum encompasses two series of autoionization resonances $5s^1np$
\cite{PhysRevA.4.2261} and $4d^9np$ \cite{Ederer:75}. These resonances
cause rapid oscillations of the time delay which are well resolved in
the present RRPA calculation but not so well in the RPAE
\cite{PhysRevA.87.063404}. As in the case of Kr, the relativistic
effects are noticeable in Xe. Similar to other atoms, the Cooper
minimum of the time delay is flattening away from the polarization
direction.

\subsection{Near threshold region}

The time delay near the threshold is dominated strongly by the Coulomb
singularity.  The scattering phase of the photoelectron propagating in
the field of the singly charged parent ion diverges to negative infinity
as the photoelectron energy goes to zero
\cite{46421597008}. Correspondingly, the photoelectron group delay
(the Wigner time delay) tends to positive infinity near threshold. The CLC
correction has a similar logarithmic singularity \cite{Dahlstrom2012}
but it is negative. So the total atomic time delay \eref{atomic} is
the sum of the two divergent terms of the opposite signs.  

\vs{5cm}
\begin{figure}[ht]
\epsfxsize=0.9\columnwidth
\epsffile{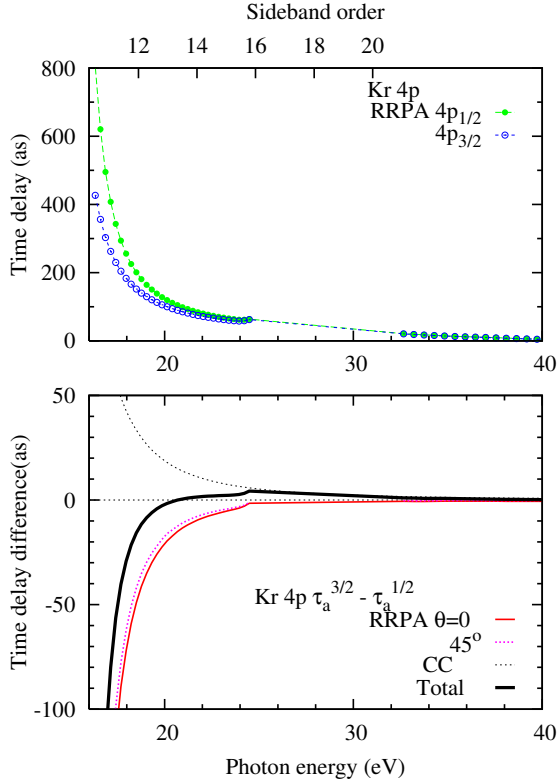}

\caption{ The atomic time delay in the Kr $4p$ subshell. Top: the Wigner
  time delay from the $4p_{1/2}$ and $4p_{3/2}$ subshells is shown with
  filled green circles and open blue circles, respectively.  Bottom:
  the time delay difference $\tau_a^{3/2}-\tau_a^{1/2}$.  The raw RRPA
  calculations at $\theta=0^\circ$ and $45^\circ$ are shown with the red
  solid and purple dashed lines, respectively. The CLC correction from
  \cite{Dahlstrom2012} is visualized with a thin dotted line. The RRPA
  result corrected by the CLC is shown with a thick solid line. 
\label{Fig4}
}
\end{figure}

\begin{figure}[ht]

\epsfxsize=0.9\columnwidth
\epsffile{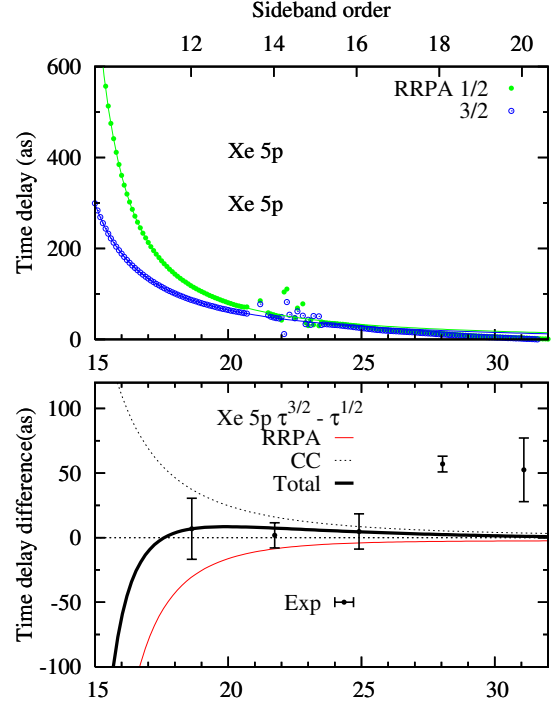}
\vs{4cm}

\caption{ Same as \Fref{Fig4} for Xe $5p$ subshell. The
  experimental data from \cite{1742-6596-635-9-092135} are plotted
  with error bars.
\label{Fig5}
}
\end{figure}

In the near threshold experiment \cite{1742-6596-635-9-092135}, it is
the difference of the atomic time delays in the $np_{1/2}$ and
$np_{3/2}$ subshells that is measured. Because of the different
ionization potentials, the photoelectron ejected from the deeper
$np_{1/2}$ subshell has a smaller kinetic energy than the one ionized
from the $np_{3/2}$ subshell. Hence the Wigner time delay in the
$np_{1/2}$ subshell is larger than the $np_{3/2}$ subshell at the same
photon energy. This characteristic behavior is seen on the top panels
of \Fref{Fig4} for Kr and \Fref{Fig5} for Xe. The same difference in
the photoelectron kinetic energies will affect the respective CLC
corrections to the time delay in the $np_{1/2}$ and $np_{3/2}$
subshells. To account for this effect, we used the values of $\tau_{\rm
CLC}$ plotted in Fig.~5 of \cite{Dahlstrom2012} as a function of
the photoelectron energy. We fitted it with an analytical formula and
calculated the difference $\tau_{\rm CLC}^{3/2}-\tau_{\rm CLC}^{1/2}$ due
to the difference in respective ionization potentials. The
corresponding difference for the Wigner time delay $\tau_{\rm
W}^{3/2}-\tau_{\rm W}^{1/2}$ was extracted from the RRPA
calculations. The area of the autoionization resonances was excluded
from this procedure because of the rapid variation of the Wigner time
delay in this region.

The resulting values of the time delay difference are shown in the
bottom panels of \Fref{Fig4} and \Fref{Fig5} for Kr and Xe,
respectively. In the case of Kr, the angular dependence of the time
delay difference $\tau_{\rm W}^{3/2}-\tau_{\rm W}^{1/2}$ is small as
can be seen in comparison of the values obtained for the two fixed
scattering angles $\theta=0^\circ$ and $\theta=45^\circ$. Hence the
values of the atomic time delay difference $\tau_a^{3/2}-\tau_a^{1/2}$
evaluated in the direction of the polarization axis can be compared
with the angular averaged experiment \cite{1742-6596-635-9-092135}.
In the case of Xe, this difference is even smaller and not noticeable
in the scale of the bottom panel of \Fref{Fig5}.

In comparison to Kr, Xe has smaller ionization thresholds. Hence, at
the same photon energy, the photoelectrons have larger kinetic energy
which take them further away from the threshold. Therefore, the effect
of the Coulomb singularity on the Wigner time delay and the CLC
correction is weaker. The net atomic time delay difference in Xe becomes
positive near threshold whereas it is negative or close to zero in the
case of Kr. These findings are in line with the experiment
\cite{1742-6596-635-9-092135}.

\section{Conclusions}

We applied the relativistic formalism and RRPA computational scheme to
evaluate the time delay in the valence shell of noble gas atoms, Ar,
Kr and Xe. The two characteristic features of the time delay are
analyzed: the angular dependence near the Cooper minimum and the
effect of the spin-orbit splitting near the threshold. Comparison with
nonrelativistic calculations serves as a convenient test and a
calibration tool.  The effect of the spin-orbit splitting is not
strong near the Cooper minimum which is relatively far away from the
ionization threshold. However, the angular dependence is significant
in this photon energy range due to efficient competition of the
nominally weak and strong photoionization channels. This dependence is
most pronounced in Ar. Indirectly, this angular dependence is
confirmed by the angular integrated experiment
\cite{0953-4075-47-24-245003} which agrees much better with angular
averaged calculations rather than angular specific data. In heavier
noble gases, Kr and Xe, the angular dependence also noticeable but not
as pronounced as in Ar.

The time delay in the near-threshold region shows little or no angular
dependence while the spin-orbit splitting effect is large. At the same
photon energy, the photoelectron ejected from a deeper $np_{1/2}$
subshell has smaller kinetic energy and less affected by the Coulomb
singularity as its counterpart ejected from the shallower $np_{3/2}$
subshell. The corresponding difference in the Wigner time delays is
offset by the difference in the correction due to the Coulomb-laser
coupling. As a result, the net atomic delay difference becomes
positive in Xe and remains negative in Kr. These findings are in line
with recent experimental observations \cite{1742-6596-635-9-092135}.

\begin{acknowledgments}
The authors wish to thank Hans Jakob W\"orner for many stimulating
discussions and Marcus Dahlstr\"om for useful comments.
ASK acknowledges support of the ARC Discovery grant DP120101805.  AM
wants to thank Dr. G. Aravind, Department of Physics, IIT Madras,
India, for very fruitful discussion.  VKD~acknowledges the support of
NSF under grant No.~PHY-$1305085$.  PCD acknowledges the support of a
grant from the Department of Science and Technology, Government of
India. STM was supported by Division of Chemical Sciences, Basic
Energy Science, Office of Science US Department of Energy.
\end{acknowledgments}



\end{document}